\documentclass[a4paper]{revtex4}

\usepackage{amsmath,amssymb}
\usepackage{epsfig}

\newcommand{\be}{\begin{equation}}
\newcommand{\ee}{\end{equation}}
\newcommand{\bi}{\begin{itemize}}
\newcommand{\ei}{\end{itemize}}
\newcommand{\ii}{\item}
\newcommand{\bea}{\begin{eqnarray}}
\newcommand{\eea}{\end{eqnarray}}
\newcommand{\gammac}{\gamma_c}
\newcommand{\f}{\frac}
\newcommand{\La}{\Lambda}
\newcommand{\cN}{{\cal N}}
\newcommand{\x}{\mathbf{x}}
\newcommand{\y}{\mathbf{y}}
\newcommand{\z}{\mathbf{z}}

\newcommand{\Lt}{\tilde L}
\def\ba{\begin{equation}}
\def\ea{\end{equation}}

\def\Bi{\begin{itemize}}

\def\Ei{\end{itemize}}

\def\ii{\item}
\def\bb{\begin{eqnarray}}

\def\nn{\nonumber \\}
\def\e{\epsilon}
\def\om{\omega}
\def\g{\gamma}
\def\d{\partial}
\def\f{\frac}
\usepackage{amsmath}
\usepackage{amssymb}
\usepackage{graphicx}

\def\abar{\bar\alpha}

\begin{document}

\title{Universality of QCD traveling-waves with running coupling}

\author{Guillaume~Beuf}
\email{gbeuf@dsm-mail.saclay.cea.fr}
\affiliation{Service de physique th{\'e}orique, CEA/Saclay,
  91191 Gif-sur-Yvette cedex, France\footnote{
URA 2306, unit\'e de recherche associ\'ee au CNRS.}}
\author{Robi Peschanski}
\email{pesch@dsm-mail.saclay.cea.fr}
\affiliation{Service de physique th{\'e}orique, CEA/Saclay,
  91191 Gif-sur-Yvette cedex, France\footnote{
URA 2306, unit\'e de recherche associ\'ee au CNRS.}}

\begin{abstract}
The  Balitsky-Kovchegov QCD equation for rapidity evolution 
describing saturation effects at high energy admits  {\it universal} 
asymptotic 
traveling-wave solutions when the nonlinear damping becomes effective. The 
asymptotic solutions   fall in universality classes depending only on some 
specific properties of the solution of the associated linear equation. We 
derive  these solutions for the recent QCD formulations of the  
Balitsky-Kovchegov equation with running coupling constant obtained from 
quark-loop calculation. While the associated linear solutions depend in 
different ways 
with 
observables  and  higher-order effects, we show that the asymptotic 
traveling-wave 
solutions all 
belong to the same universality class whose solutions are given. Hence the 
influence 
of saturation  
stabilizes the QCD evolution with respect to higher order effects and 
leads to universal features at high enough rapidity, such as the form of the 
traveling waves, the intercept of the saturation scale and geometric scaling 
in  square-root of the rapidity. 
\end{abstract}

\maketitle

%%%%%%%%%%%%%%%%%%%%%%%%%%%%%%%%%%%%%%%%%%%%%%%%%%%%%%%%%%%%%%%%%%%%%%%%%%%%
%%%%%%%%%%%%%%%%%%%%%%%%%%%%%%%%%%%%%%%%%%%%%%%%%%%%%%%%%%%%%%%%%%%%%%%%%%%%
%%%%%%%%%%%%%%%%%%%%%%%%%%%%%%%%%%%%%%%%%%%%%%%%%%%%%%%%%%%%%%%%%%%%%%%%%%%%

\section{Introduction}
\label{sec:intro}

In the  leading-logarithm (LL) approximation of high-energy 
(high-density) QCD, the evolution with rapidity 
$Y$ of deep-inelastic scattering cross-sections is driven by the nonlinear 
Balitsky-Kovchegov 
(BK) equation \cite{Balitsky:1995ub, Kovchegov:1999yj, 
Kovchegov:1999ua}. This equation is supposed to capture essential features 
of 
saturation effects in the ``mean-field'' approximation where fluctuations 
(or 
``Pomeron-loop'' effects \cite{loops}) can be neglected. More specifically, 
for 
the  dipole-target amplitude in 2-dimensional transverse position space 
$\cN(\x,\y,Y),$ 
it reads
\begin{equation}
\label{eq:bk}
\partial_Y \cN(\x,\y,Y) = \frac{\bar{\alpha}}{2\pi}\int d^2 z \ 
\frac{\vert xy\vert ^2}{\vert xz\vert ^2\vert zy\vert ^2} 
\left\{\cN^{}(\x,\z,Y)+\cN(\z,\y,Y)-\cN(\x,\y,Y)-\cN(\x,\z,Y)\times\cN(\z,\y,
Y) 
\right\}\ 
,
\end{equation}
where $\vert xy\vert ^2 
\equiv (\x\!-\!\y)^2,$  $\x$ and $\y$ are the transverse space positions of 
the quark and 
antiquark constituting the QCD dipole. In Eq.\eqref{eq:bk}, the coupling 
$\bar\alpha = 
\alpha_s 
N_c/\pi$ is kept
fixed, since the derivation of the equation is made at leading logaritmic 
approximation, and thus zeroeth perturbative order in the coupling. Indeed, 
this 
nonlinear equation corresponds to 
resumming QCD ``fan diagrams'' in the LL 
approximation \cite{Kovchegov:1999yj,Kovchegov:1999ua}. 

In the 1-dimensional case, without
impact-parameter dependence (\emph{e.g.} for large targets), 
$\cN(\x,\y)\equiv 
\cN (\vert xy\vert )$, the equation
acquires 
a 
simple 
form in the Fourier-transformed momentum space, namely 
\cite{Kovchegov:1999ua}
\be
{\bar\alpha}^{-1} \partial_Y N (L,Y) = 
\, \chi (-\partial_L)\: N (L,Y) - \, N^{\,2} (L,Y),
\label{1}
\ee
where $L(k) = \log (k^2/\Lambda^2),$ and  $\Lambda$ is here an 
arbitrary 
scale. The Fourier-transformed amplitude
\be
\label{fourier}
N(L,Y)=\int_0^{\infty} \frac{d\vert xy\vert }{\vert xy\vert }\ J_0(k\vert 
xy\vert 
)\,\cN(\vert xy\vert ,Y)
\ee 
can be related by convolution (see \emph{e.g.} \cite{Bialas:2000xs}) to the 
{\it 
unintegrated} gluon distribution in the 1-dimensional space in
transverse momentum $k\equiv \vert {\bf k}\vert .$ 
In this LL approximation the characteristic function of the kernel has 
the standard  Balitsky-Fadin-Kuraev-Lipatov (BFKL) form
\cite{Lipatov:1976zz, Kuraev:1977fs, Balitsky:1978ic}, namely 
\be
\label{eq:bfkl_ll}
\chi(\gamma) = 2\psi(1) - \psi(\gamma) - \psi(1-\gamma)\ .
\ee
Quite recently, the extension of the BK equation to running coupling has 
been 
the subject of interesting theoretical studies. More generally, the 
extension 
of the saturation formalism beyond LL order is of present 
concern.

The goal of the present paper is to apply the traveling-wave method  to the 
problem of solving the nonlinear BK equation with running coupling. This 
method 
allows to find  asymptotic solutions to nonlinear equations of the BK type 
and 
to 
discuss their 
universality properties. 

Indeed, we will 
consider the recent  theoretical advances towards a full QCD formulation 
of 
the BK equation at next-to-leading logarithmic (NLL) order in the coupling 
constant 
and beyond. This implies in particular taking into account in a proper way 
the 
running QCD coupling and NLL and higher order contributions to the kernel.
The first part of our study is devoted to the equations 
incorporating the  terms determined through the quark-loop  
contributions to the BK equation
\cite{Balitsky:2006wa,Kovchegov:2006vj,Kovchegov:2006wf} (see also related 
older \cite{Levin:1994di} and  recent 
\cite{Gardi:2006rp,Fadin:2006ha,Fadin:2007ha} references). Then, we 
 consider the full NLL contributions to the kernel 
\cite{Fadin:1998py,Ciafaloni:1998gs} and the higher-order contributions 
implied 
by the renormalization-group constraints 
\cite{Salam:1998tj,Ciafaloni:1998iv,Ciafaloni:1999yw,Ciafaloni:2003rd}.

Since a few years, the solutions of saturation with running coupling have been 
considered in the literature. 
Already in the papers \cite{Gribov:1984tu,Iancu:2002tr}, the dominant 
asymptotic term of the 
solution has been obtained from the knowledge of the BFKL equation. Using 
this equation with absorbing boundary conditions, the subasymptotic term 
could also be derived \cite{Mueller:2002zm} and extended to NLL BFKL kernels 
\cite{Triantafyllopoulos:2002nz}.

Our tools for the analysis rely on the powerful traveling-wave method. 
Indeed, 
it 
has been remarked that, contrary 
perhaps to naive expectation, the nonlinear character of the BK equation 
leads 
to some simplification w.r.t. the  linear BFKL equations. Interesting {\it 
universal} properties, \emph{i.e.} asymptotic 
solutions which do not depend either on initial conditions, on the precise 
form 
of the nonlinear terms or on details of the linear kernel.
 This is tightly linked to a 
mathematical property of a well-known class of nonlinear equations 
\cite{Fisher:1937,Kolmogorov:1937} which admits asymptotic solutions in 
terms 
of 
{\it traveling-waves} \cite{Bramson,Brunet:1997,ebert}. The relevance of 
this class of equations and traveling-waves for the BK equation has been 
raised 
in 
\cite{Munier:2003vc,Munier:2003sj}.

There exist  previous analytical results  relevant for our study on the  BK 
equation with 
running 
coupling.
Heuristically, the running coupling has been considered together 
with the 1-dimensional BK 
equation \eqref{1} by substituting the fixed coupling in front of the 
equation with a 
running coupling $\abar(L)$ depending on the {\it gluon} momentum scale 
\cite{Mueller:2002zm,Munier:2003sj}. One writes
\be
\label{alpha(L)}
{\bar\alpha(L)}^{-1} \partial_Y N (L, Y) = 
\chi (-\partial_L)\, N (L, Y) - N^{\, 2} (L, 
Y)\ .
\ee
We define $L(k) = \log (k^2/\Lambda^2),$ where  $\Lambda$ is now the QCD scale. 
One 
considers the LL kernel function as in 
Eq.\eqref{eq:bfkl_ll} and the running coupling reads
\be
\label{eq:b_def}
\bar\alpha(L) = \f{1}{bL}\ ,
\qquad \qquad
b = \f{11 N_c - 2 N_f}{12 N_c}\ .
\ee 
In this case, the asymptotic universal solution has been found, either by 
approximation of the linear regime with absorbing conditions 
\cite{Mueller:2002zm,Triantafyllopoulos:2002nz} or directly 
\cite{Munier:2003vc,Munier:2003sj} from the 
traveling 
wave method of solving equation \eqref{alpha(L)}.

In Ref.\cite{Balitsky:2006wa}, the quark-loop contributions to the small-$x$ 
evolution  has been computed for the dipole-target amplitude in 
transverse-position space. Through a   separation scheme 
between the running coupling at next-to-leading order level and the kernel at 
NLL 
level, an 
evolution equation  has been 
derived. 
 For the  1-dimensional amplitude $\cN(\vert 
xy\vert,Y)$ 
the 
correspondingly modified  BK  equation (see \cite{Balitsky:2006wa}, 
formula (52) where the equation is in fact written 
using traces over Wilson-line operators,  easily translated in 
terms 
of $\cN$) reads
\bb
b \tilde{L} \  \f{\d }{\d Y} \mathcal{N}(\vert xy\vert,Y) =   \int\f{d^2 
{z}}{2 
\pi} 
\left[ \mathcal{N}(\vert {xz}\vert ,Y) +\mathcal{N}(\vert {zy}\vert ,Y) 
-\mathcal{N}(\vert xy\vert,Y)-\mathcal{N}(\vert xz\vert ,Y)\ 
\mathcal{N}(\vert 
zy\vert ,Y) \right]\ &\times& \nn
 \times \left\{\f{\vert xy\vert^2}{\vert xz\vert ^2\ \vert zy\vert ^2} 
+\f{1}{\vert xz\vert ^2}\left(\f{\abar(\vert xz\vert)}{\abar(\vert 
zy\vert)}-1\right)+\f{1}{\vert yz\vert ^2}
\left(\f{\abar(\vert yz\vert)}{\abar(\vert zx\vert)}-1\right)\right\} \ 
&,&\label{balitsky1}
\eea
where
\be
\label{eq:Ltilde}
\abar(\vert xy\vert) = \f{1}{b \tilde{L}(\vert xy\vert)}\ ,
\qquad \qquad
\tilde{L}(\vert xy\vert)=- \log(\vert xy\vert ^2 \Lambda^2)\ .
\ea Note 
that 
the expression \eqref{balitsky1} needs some regularization procedure, since 
Landau 
poles appear in the $z$-integration over $\abar(\vert xz\vert)$ and  
$\abar(\vert 
yz\vert).$ We will consider this important point in  detail later on for our 
study.

In the same time, another  consistent derivation of the BK equation with 
running 
coupling constant has been performed both for the dipole amplitude in 
position 
space and for the 
unintegrated gluon distribution in momentum 
space \cite{Kovchegov:2006vj,Kovchegov:2006wf}. The 
same quark-loop contribution as in \cite{Balitsky:2006wa} has been found, but 
using a different separation scheme. The quark-loop contributions have 
been 
splitted differently into a contribution to the running coupling and a 
contribution to the  kernel. The 
equation found in position space is thus different from \eqref{balitsky1}. 
The 
involved running couplings 
appear to form a ``triumvirate'' structure in terms of three different 
scales. 
More precisely (see 
\cite{Kovchegov:2006wf}, where
formula (48) the equation is written 
with additional factors $e^{-5/3}$  absorbed here in 
$\La$) one 
writes
\bb
\f{\d \tilde{\phi}}{\d Y} (k,Y) &=&  \int \f{d^2 \bf{q}}{2 \pi}\ \f{\abar 
(q^2) \ 
\abar (({\scriptstyle{\bf k\!-\!q}})^2)}{\abar (k^2)}
 \left[\f{1}{({\scriptstyle{\bf k\!-\!q}})^2}\ \tilde{\phi}(q,Y) + 
\f{1}{q^2}\ 
\tilde{\phi}(\vert {\scriptstyle  {\bf k\!-\! q}}\vert ,Y)
 - \f{k^2}{q^2 ({\scriptstyle {\bf k\!-\!q}})^2}\  \tilde{\phi}(k,Y) \right] 
+ 
nonlinear \ terms
\label{EqTrium}
\eea
with  
\ba
\tilde{\phi}(k,Y)= {\abar(k^2)}^{-1}{\phi(k,Y)} \ ,
\label{phi}
\ea
$\phi(k,Y)$ being the unintegrated gluon distribution in tranverse 
momentum 
space. 
The {\it nonlinear terms} in Eq.\eqref{EqTrium} are not explicitely written, 
but as already known from other studies of 
the BK 
equation in momentum space \cite{marquet}, their precise form is not needed 
for 
the derivation of the traveling-wave solutions in the transition region 
towards 
saturation, provided they ensure the necessary damping.

This ``triumvirate'' of coupling constants has been advocated  long ago from 
bootstrap properties  \cite{Levin:1994di}, while the definition \eqref{phi} 
has 
been introduced more recently \cite{Kovchegov:2006wf}. Quite interestingly,  
the 
consistency of the equation for the dipole amplitude 
\eqref{balitsky1} with the one for the unintegrated gluon distribution has 
been established in \cite{Kovchegov:2006wf}.
However, 
the corresponding Pomeron intercept driving the leading rapidity 
behaviour   
of the linear equation is different in both cases at NLL order, by contrast 
with the 
identical rapidity behaviour at LL level. The question 
of the 
universality properties of 
the 
asymptotic solutions of these  equations 
(\ref{alpha(L)},\ref{balitsky1},\ref{EqTrium}) when taking into account 
nonlinear 
damping terms is thus interesting to investigate. Indeed, the dependence on 
the 
 separation schemes and the dependence  of the Pomeron intercept on the 
observable valid for the linear regime at NLL order and beyond have to be 
reexamined when including saturation effects.

Concerning the  scheme dependence related to the renormalisation-group 
constraints, one starts with the expectation that 
the  linear regime of the BK equation is  
driven 
by the full NLL  BFKL kernel 
already 
available from the calculation of \cite{Fadin:1998py,Ciafaloni:1998gs}. It is 
known that one introduces a ``renormalization-group improved'' (RG-improved) 
kernel at all 
orders of perturbation. One has   to get rid of unwanted spurious 
singularities 
brought together with the NLL calculation. This boils down  to the existence 
of 
different  RG-improved schemes which are equivalent at NLL accuracy but 
with 
different 
resummations at higher orders (see, \emph{e.g.},  the S3 and S4 schemes 
\cite{Salam:1998tj} and the CCS scheme \cite{Ciafaloni:1999yw}). 

The  traveling-wave method for the BK equation with the NLL and higher-order 
kernels have already been used in 
Refs.\cite{Enberg:2006aq,Peschanski:2006bm}. 
In Ref.\cite{Enberg:2006aq}, the NLL kernels have been  considered together 
with a fixed value of the QCD coupling. In this case also the method works 
for 
finding suitable asymptotic solutions, and they depend on both the value of 
the 
coupling and on the NLL framework. In some sense, our study is a continuation 
of  
Ref.\cite{Peschanski:2006bm} where the  RG-improved 
scheme dependence is studied for the heuristic equation \eqref{alpha(L)}.  
Note 
that  using the fixed coupling 
in 
the  definition of the  RG-improved scheme leads to a scheme dependence 
of the traveling waves \cite{Peschanski:2006bm}. We want to reanalyze the non linear problem when 
using a 
running coupling 
for the scheme definition, as done for the corresponding linear problem in  
\cite{Ciafaloni:1999yw,Ciafaloni:2003rd}.

It is important to note that, for the Balitsky scheme, Eq.\eqref{balitsky1}, as 
well as 
for the ``triumvirate case'', Eq.\eqref{EqTrium}, the question of spurious 
singularities 
and of  renormalization-group improved kernels have not yet been studied. 
Moreover, the regularization of Landau poles may introduce some extra 
subtelties in this problem. %%
We will show, by a careful study of  singularities
 in the NLL kernels, that such collinear singularities do 
not change 
the determination of  the universal 
terms in the asymptotic expansion but they are responsible of a slowing down of the convergence of 
the asymptotic expansion and probably  introduce stronger 
nonuniversal 
subasymptotic contributions. Such subasymptotic contributions have been 
discussed   \emph{e.g.} in Ref.\cite{Triantafyllopoulos:2002nz}. They 
certainly deserve a separate further
study  in the framework of traveling waves.

%%%%%%%%%%%%%%%%%%%%%%%%%%%%%%%%%%%%%%%%%%%%%%%%%%%%%%%%%%%%%%%%%%%%%%%%%%%%

The plan of the paper is as follows. In section \ref{sec:alpha(L)} we 
consider  
the BK 
equation \eqref{alpha(L)} with  momentum-space running coupling and recall 
the 
known  results on
 its  
traveling-wave  solutions \cite{Munier:2003sj,Peschanski:2006bm},  including %%
 a new discussion of the
 effect of NLL kernel singularities on the universality properties.%% 
 In  section 
 \ref{sec:alpha(r)} we turn to 
position 
space and 
derive the corresponding  linear-regime and traveling-wave solutions 
for the Balitsky scheme of \cite{Balitsky:2006wa}. In section  
\ref{Triumvirate}
we consider the ``triumvirate ''  case 
\cite{Kovchegov:2006vj,Kovchegov:2006wf}, 
and exhibits its  
solutions 
in the same way. In section  \ref{Scheme} we discuss the  RG-improved scheme 
dependence. In section  \ref{Conclusion} we summarize our results by 
concluding 
that 
all corresponding traveling-wave solutions fall into the same universality 
class and indicate the corresponding asymptotic predictions.

%%%%%%%%%%%%%%%%%%%%%%%%%%%%%%%%%%%%%%%%%%%%%%%%%%%%%%%%%%%%%%%%%%%%%%%%%%%%
%%%%%%%%%%%%%%%%%%%%%%%%%%%%%%%%%%%%%%%%%%%%%%%%%%%%%%%%%%%%%%%%%%%%%%%%%%%%
%%%%%%%%%%%%%%%%%%%%%%%%%%%%%%%%%%%%%%%%%%%%%%%%%%%%%%%%%%%%%%%%%%%%%%%%%%%%

\section{BK equation with running coupling constant in momentum-space}
\label{sec:alpha(L)}
\subsection{Solution of the linear equation}
Let us first remind briefly the method used in 
Ref.\cite{Munier:2003sj} to obtain the traveling-wave solutions  of the 
Balitsky equation
\eqref{alpha(L)}.
Following the general method \cite{Munier:2003sj}, we first write the 
solution 
to the linearized version of the 
equation. In the saddle-point approximation, it has the form  of a double 
Mellin 
transform~\cite{Ciafaloni:1999yw}
\be
\label{eq:N}
N(L,Y) = \int \frac{d\gamma}{2\pi i}
         \int \frac{d\omega}{2\pi i} \,
         N_0(\gamma,\omega) \,
         \exp\left(-\gamma L + \omega Y + \frac{1}{b\omega} X(\gamma)
            \right),
\ee
with the kernel dependence appearing through the function
\be
\label{eq:capital_x}
X(\gamma) = \int^{\gamma}_{\hat{\gamma}} 
                   d\gamma' \, \chi (\gamma')\ ,
\ee
$\hat\gamma$ being an unspecified constant. Indeed, using the 
saddle-point 
method for the integration over $\gamma$ at large enough $L,$ one gets the  
equation
\be
\label{eq:saddle}
-L+\frac{1}{b\omega}\ \partial_{\g} X(\gamma)\equiv -L+\frac{1}{b\omega}\ 
\chi(\gamma) = 0\ ,
\ee
or equivalently in the form of an operator acting on the amplitude
\be
\label{eq:consistency}
bL\, \partial_Y N (L, Y) = 
\chi (-\partial_L)\, N (L, Y)\ .
\ee
Hence the solution \eqref{eq:N} verifies the restriction to linear terms of 
\eqref{alpha(L)}, within the saddle-point approximation.

As a next step one performs 
the saddle point integration over $\omega$ in the limit of large $Y$.
The saddle point $\omega_s$ is given by 
\be
\label{l}\omega_s = 
\sqrt{
\frac{ X(\gamma) }{bY}}\ ,
\ee
behaving like 
$\omega_s \sim Y^{-\frac{1}{2}}$. With this form of $\omega_s$ the gluon 
density is given by
\be
\label{eq:ampl_int}
N(L,Y)  \sim \int \frac{d\gamma}{2\pi i}\ N_0(\gamma)  
\ \exp(-\gamma L + \Omega(\gamma) t),
\ee
where the time variable is interpreted as $
t = \sqrt{Y}$ and 
\be
\label{eq:disp_rel}
\Omega(\gamma) = 
\sqrt{\frac{4}{b} X(\gamma) }\ .
\ee
\subsection{Traveling-wave solutions}
A  critical 
{\it group 
velocity} 
(defined as the minimum of the {\it phase velocity} in the wave language) is 
obtained as 
\be
\label{eq}
\upsilon_g = 
\Omega(\gamma_c)/\gamma_c = \Omega'(\gamma_c)\ .
\ee
However, $\gammac$ determined in such a way still depends on the arbitrary 
constant $\hat\gamma$.
Thus,  requiring $\upsilon_g$ to be independent on the choice of 
$\hat\gamma$  
means 
$d\upsilon_g(\hat\gamma)/d\hat\gamma = 0,$ which in turn gives  
$d\upsilon_g(\gamma_c)/d\gamma_c=0$ since the dependence of the velocity on 
$\hat\gamma$ comes through 
$\gamma_c$ only. 
Applying this condition to Eq.~\eqref{eq:disp_rel} one   gets
\be
\label{criticalbis}
d\upsilon_g(\hat\gamma)/d\hat\gamma 
=\f{d\left(\Omega(\gamma_c)/\gamma_c\right)}{d\gamma_c}\ = 0 
\quad \Rightarrow \quad 
\upsilon_g\  =\ 
\sqrt{\frac{2\chi(\gammac)}{b\, \gammac}}\ ,
\ee
eliminating all dependence on the arbitrary constant $\hat \g$ in the 
definition \eqref{eq:capital_x}. As a consequence, one finds the equation for 
the 
critical 
exponent
\be
\gammac = \f{\chi(\gammac)}{\chi'(\gammac)}\label{saddle}\ ,
\ee
which is the well-known value ($\gammac = .6275$) associated to the 
saturation 
solutions using the  LL BFKL kernel \eqref{eq:bfkl_ll}.

Finally, using an ansatz technique borrowed from statistical physics 
\cite{Brunet:1997}, one finds  \cite{Munier:2003sj} the result for the gluon 
density 
\be
 N(L,t) = 
{\rm const} \cdot t^{\frac{1}{3}} \cdot {\rm Ai} \left(
\left(\frac{\sqrt{2 \gammac b \, 
\chi(\gammac)}}{\chi''(\gammac)}\right)^{\frac{1}{3}}
\log\frac{k^2}{Q^2_s(t)} \; t^{-\frac{1}{3}} +
\xi_1 \right) \cdot
\left(\frac{k^2}{Q^2_s(t)}\right)^{-\gammac},
\label{gluon}\ee
in agreement with \cite{Mueller:2002zm}. $\xi_1 = -2.338$ is the first zero of 
the Airy function ${\rm 
Ai}(\xi)$
and the saturation scale has the form
\be
Q^2_s(t) = 
Q_0^2\, \exp \left(
\sqrt{\frac{2 \chi(\gammac)}{b \gammac}}\; t + 
\frac{3}{4} 
\left(\frac{\chi''(\gammac)}{\sqrt{2 \gammac b \, 
\chi(\gammac)}}\right)^{\frac{1}{3}}\xi_1\; t^{\frac{1}{3}}\right)\ ,
\label{satur}\ee
 up to a non universal multiplicative constant. Hence the two first terms of 
the 
expansion a large $t$ of ${d \log(Q^2_s)}/{dt}$ (and thus of the 
{\it saturation intercept} ${d \log(Q^2_s)}/{dY}$) are completely specified. 
Note 
that the result \eqref{gluon} is known to be valid in the transition region 
towards saturation but not in the deep infrared saturation domain. However,  
the 
infra-red region  is also universally constrained by unitarity, namely 
$N \sim 
\log (Qs/k)$.
 
 The important output of this traveling-wave solution with running coupling 
constant 
$\alpha(L)$ is that the amplitude \eqref{gluon} depends only on the ratio 
${k^2}/{Q^2_s(t)}$ at asymptotic values 
of 
$t.$  It thus corresponds to a 
geometric-scaling property \cite{geom} of the gluon 
density as a function of $t\equiv \sqrt Y$ \cite{Gelis:2006bs}, instead of 
$Y$ in
the non-running 
case. 

It is useful to note that the ansatz leads to the  solution 
(\ref{gluon},\ref{satur}) 
which do not depend on the saddle-point approximation used initially in  
\eqref{eq:N}. Indeed, all parameters of the ansatz are determined 
self-consistently 
from the asymptotic analysis of the original nonlinear equation. The 
saddle-point 
approximation has been useful to determine the appropriate form of the ansatz 
\cite{Munier:2003sj}.

%%%%%%%%%%%%%%%%%%%%%%%%%%%%%%%%%%%%%%%%%%%%%%%%%%%%%%%%%%%%%%%%%%%%%%%%%%%%

\subsection{Extension of traveling waves to a $\omega$-dependent kernel}
\label{sec:omega}

In the following, and throughout the rest of the paper,  we introduce the 
possibility of dealing with a more general kernel $\chi(-\d_L, \d_Y).$ This 
generalization comes naturally from the NLL studies. For instance it has 
been introduced in the
$\omega$-expansion method of Ref.\cite{Ciafaloni:1998iv}, in order to define 
RG-improved kernels. More generally, it is useful in order to 
perform a change of variable $\abar (L) \to \omega$ in the linearized equation, 
which is the starting point of the traveling wave method. It allows to 
express the solution of the linear equation in terms of a linear 
superposition of waves. The nonlinear term will then select the critical 
wave as the traveling wave solution.

Let us recall for further use the extension \cite{Peschanski:2006bm} of 
the traveling wave method in 
the case of a kernel $\chi(-\d_L, \d_Y)$ in \eqref{alpha(L)}, 
namely
\be
\label{alpha(L)w}
{\bar\alpha(L)}^{-1} \partial_Y N (L, Y) = 
\chi(-\d_L, \d_Y)\, N (L, Y) - N^{\, 2} (L, 
Y)\ .
\ee
In double Mellin space, the linear part of the equation  reads 
$b L \omega = \chi(\g, \omega)$. Introducing 
\be
\label{eq:capital_xo}
X(\gamma, \omega) = \int^{\gamma}_{\hat{\gamma}} 
                   d\gamma' \, \chi (\gamma',\omega)\ ,
\ee the saddle-point integration over $\omega$ leads to  the asymptotic 
solution in $Y$ 
\ba N(L,Y)=\int \f{d\g}{2\pi i} N_0(\g) \ \exp \left[ -\g L + 
\f{1}{b \omega_s} \left( 2 X(\g,\omega_s)-\omega_s \dot{X}(\g,\omega_s) 
\right)\right] \label{saddleresult} \ ,\ea
where $\omega_s$ is given by the saddle-point equation 
\ba \label{eq:cond2}
 Y b \omega_s^2-X(\g,\omega_s)+\omega_s \dot{X}(\g,\omega_s)=0  \ea
and the ``dot'' means the derivative with 
respect to 
$\omega.$

The solution $\om_s(\g,Y)$ of Eq.\eqref{eq:cond2} defines the asymptotic 
properties of the amplitude \eqref{saddleresult}. In order to solve it, one has 
to take into account the analytic properties of the kernel integral 
\eqref{eq:capital_xo}, which posesses singularities near $\g=0,1.$ The solution 
 will depend on the position of the critical value $\g=\g_c$ for the saturation 
solution with respect to the singularities. Indeed, as we shall now prove, one 
has to avoid some well-defined regions around the  singularities in order to find 
the universality properties. 

Let us first consider the situation where $\g_c$ is not in the neighbourhood of the 
singularities of the kernel. Then, the  
expansion of  the integral of the kernel \eqref {eq:capital_xo} 
and its derivative at $\omega =0$ is analytic 
 near $\g \sim \g_c.$ One can use the Taylor expansion around  $\omega =0$
\be
\label{eq:kernel_expansion}
X\,(\gamma, \omega)  =
 \sum^{\infty}_{p=0} \frac{X^{(p)}(\gamma,0)}{p!}\ \omega^{p}\ ,
 \ee
 and by collecting  terms with the same powers of $\omega_s,$ Eq.~(\ref{eq:cond2}) 
writes 
\be
\label{sum}
\left[Yb + \frac{1}{2} \ddot{X}\, (\gamma, 0) \right] \omega^2_s =
X\, (\gamma, 0) - 
\left\{\sum^{\infty}_{p = 3} \frac{1}{p(p-2)!} X^{(p)}\, (\gamma, 0)\, 
\omega^p_s 
\right\}\ .
\ee
 Following 
\cite{Peschanski:2006bm}, the term in braces will contribute only to 
subleading 
nonuniversal terms in the traveling wave expansion while the left-hand term 
with the second derivative w.r.t. $\omega$ can be 
absorbed by a translation in $Y.$ This term (and the higher derivative terms) 
depends thus on initial conditions, and we have the universal asymptotic 
behaviour at large rapidity
\be
\label{omega}
\omega_s = \sqrt{\f{ X\, (\gamma, 0)}{bY}} .
\ee
Hence the previous relations (\ref{l}-\ref{saddle}) remains valid once 
substituting $\chi(\g) \to \chi(\g,\om=0).$

Now, we have to take into account the singularity regions,
 in order to discuss the validity range
 of the previous derivation.
The singularities of the function $X(\g,\omega)$  depend crucially 
of the use of  a renormalization-group improved kernel or not, due to 
the presence 
of spurious singularities in NLL contributions of the kernel, which have 
to be cancelled by the  renormalization-group improvement scheme.
 We shall however 
consider also the case when spurious singularities are still present, by 
comparison.

\begin{itemize}
\item {\it Renormalization-group improved kernels.} The  
behaviour of the kernels $\chi$ near the singularities are simple poles (apart 
possibly from  the quark contributions, which we shall not consider in this 
derivation). By integration, see \eqref{eq:capital_xo}, 
this leads to logarithmic singularities in  the 
function $X(\g,\omega)$ and thus in Eq.\eqref{eq:cond2}. Hence, the singularity 
region  to be avoided is $\g_c , 1\!-\!\g_c \le {\cal O}
\left(\om_s \log|\om_s|
\right)
=  {\cal O}\left({\log Y}/{\sqrt{Y}}\right).$  
Inside this region, the expansion \eqref{sum} and thus the universal hierarchy 
in $\om_s$ is spoiled.

\item {\it NLL kernels with spurious singularities.} The  
 NLL kernels $\chi$ contain  singularities up to triple poles. 
By the integration in \eqref{eq:capital_xo}, new single and 
double-pole singularities appear at next-to-leading 
order. Hence, the singularity region to be avoided becomes  
$\g_c, 1\!-\!\g_c \le  {\cal O}\left(\om_s^{1/2}
\right)
= {\cal 
O}\left({Y}^{-1/4}\right).$  Inside this region, the expansion \eqref{sum} and 
thus the universal hierarchy in $\om_s$ is spoiled. Thus the useful region 
of analyticity requires a much higher rapidity to be valid.
\end{itemize}

This discussion shows that the universality properties will show up 
at different scales in rapidity, depending on the singularity structure 
of the NLL kernels. In fact, if one considers   NLL contributions 
keeping spurious singularities, subasymptotic nonuniversal terms 
may be dominant up to high rapidities. Hence, the singularity structure 
of the NLL kernels will affect the subasymptotic behaviour.

We note that the running coupling case, introducing the integration  
\eqref{eq:capital_xo}, softens the singularities. It may also happen that the 
$\om$-expansion itself   softens the singularities (even the spurious 
ones, if present), hence possibly improving the rapidity range of universality 
properties. This, and the question of quark contributions deserve a more 
detailed study. In the present work we will focus on the case where the 
critical $\g_c$ is well inside the analyticity region. Since $\g_c$  takes 
the same universal value  $\g_c = .6275$ as for the LL kernel, the universality 
properties will appear when the region of singularities is avoided and thus for 
large enough rapidities. The price to pay for the eventual presence of spurious 
singularities is that the convergence to the universal features may be sizeably 
pushed away in rapidity. %% 

In order to finally get the universal asymptotic terms of traveling-wave 
solutions, one has to express the linearized version of the BK equation 
\eqref{alpha(L)w} retaining only terms in the kernel relevant for the 
asymptotc 
analysis. By
expanding around $\omega=0$, $\g=\g_c$ up to the second order, one gets
\bb
\label{eq:bk_Lt}
\frac{bL}{2t}\, \partial_t N 
& = &  \left\{-\frac{b}{2}\upsilon_g^2\, \partial_L + 
                        \frac{1}{2}\chi''(\, \partial_L^2 +
                        2 \gammac \, \partial_L +
                        \gammac^2) \right.
\nonumber \\
&   & \left. \ + \
      \frac{1}{2t} \, \dot{\chi}\, \partial_t  - \frac{1}{2t} \, 
\dot{\chi}'\, 
\partial_L \partial_t - \frac{1}{2t} \, \dot{\chi}'\, \gammac \partial_t +
      \frac{1}{8t^2} \,\ddot{\chi}\, (\partial_t^2 - \frac{1}{t}\, 
\partial_t)\right\}N \ ,
\eea
 where the ``prime'' is the derivative with respect to 
$\gamma$.

The terms in the first line of \eqref{eq:bk_Lt} correspond to the 
same expansion as for the $\om$-independent  case and they  contribute up to 
order  $Y^{-1/6},$ which is enough to determine the two first asymptotic 
terms 
of 
the amplitude and the saturation intercept. The second line 
contains new terms, with derivatives in $\om$, which contribute only to 
higher 
order. 

It is interesting to note however that the first dominant term in the 
second line of  \eqref{eq:bk_Lt} contributes with order  $Y^{-1/3}.$ The 
nonuniversal contributions are expected to start at  order $Y^{-1/2},$ since 
they 
correspond to  a shift  $t\to 
t+t_0$ which results in nonuniversal terms of order $Y^{-1/2}.$ Hence 
the first  term  in the 
second line of  \eqref{eq:bk_Lt}
gives a  contribution related to the  $\omega$-dependence of the kernel, since 
it 
depends on $\dot \chi$ which could be the remaining track of NLL effects in  
the 
universal traveling-wave solutions.

All in all,  one finds the same solutions as 
(\ref{gluon},\ref{satur}) with the substitution  $\chi(\g) \to 
\chi(\g,\om=0).$

%%%%%%%%%%%%%%%%%%%%%%%%%%%%%%%%%%%%%%%%%%%%%%%%%%%%%%%%%%%%%%%%%%%%%%%%%%%%
%%%%%%%%%%%%%%%%%%%%%%%%%%%%%%%%%%%%%%%%%%%%%%%%%%%%%%%%%%%%%%%%%%%%%%%%%%%%
%%%%%%%%%%%%%%%%%%%%%%%%%%%%%%%%%%%%%%%%%%%%%%%%%%%%%%%%%%%%%%%%%%%%%%%%%%%%

\section{BK equation with running coupling for Balitsky's scheme}
\label{sec:alpha(r)}

%%%%%%%%%%%%%%%%%%%%%%%%%%%%%%%%%%%%%%%%%%%%%%%%%%%%%%%%%%%%%%%%%%%%%%%%%%%%

\subsection{Balitsky's formalism and regularization}

Let us now derive the traveling-wave solutions for Eq.\eqref{balitsky1}. 
Following 
the abovementionned method, one first look for the solution of its linear 
part. 
However, in order to give  mathematical and physical consistency, one has 
to 
introduce some regularization procedure of the integration over $z$ in 
position 
space in order to avoid the Landau poles at $\Lt (\vert xz \vert)=0$ and $\Lt 
(\vert yz \vert)=0$. 

Starting  with solving the linear part of \eqref{balitsky1} 
\bb
b \tilde{L} \  \f{\d }{\d Y} \mathcal{N}(\vert x y \vert,Y) &=&  
\int_{\cal{R}}
 \f{d^2 z}{2 \pi} 
\left[ \mathcal{N}(\vert  xz\vert ,Y) +\mathcal{N}(\vert z y \vert ,Y) 
-\mathcal{N}(\vert x y \vert,Y)\right]\ 
\times \nn
&\times&\left\{\f{{ \vert x y \vert}^2}{\vert  xz\vert^2 \vert z y 
\vert^2} 
+\f{1}{\vert  xz\vert^2} 
\f{\log \left({\vert z y \vert^2}/{\vert  xz\vert^2}\right)}{\log (\vert  
xz\vert^2
 \Lambda^2)}+\f{1}{\vert z y \vert^2}\f {\log \left({\vert  xz\vert^2}/{\vert 
z y 
\vert^2}\right)}{\log(\vert z y \vert^2 
\Lambda^2)}\right\}  \ ,\label{balitsky2}
\eea
where the subscript ${\cal{R}}$ denotes the regularization procedure, we 
insert 
the Mellin transforms $\hat{\mathcal{N}}(\gamma, Y)$ of the 
dipole amplitude in position space.
\ba \mathcal{N}(r,Y) \equiv \int \f{d \gamma}{2 \pi i} \, e^{- \gamma 
\tilde{L}(r)} \, 
\hat{\mathcal{N}}(\gamma, Y)
= \int \f{d \gamma}{2 \pi i} \int \f{d \omega}{2 \pi i} \, e^{\omega Y - 
\gamma 
\tilde{L}(r)} \, \tilde{\mathcal{N}}(\gamma, \omega)\ . \label{Melr}
\ea
Applying the Mellin-transform \eqref{Melr} to  the right-hand side of 
\eqref{balitsky2}, one  obtains the diagonalized 
form of
 the linear kernel
\ba
\chi^{Bal}_{\cal{R}}(\g,\abar(\vert  xy\vert))  =  
\int_{\cal{R}} \f{d^2 z}{2 \pi} 
\left[\left(\f{\vert xz\vert}{\vert xy\vert}\right)^{2 \g} +\left(\f{\vert 
yz\vert}{\vert xy\vert}\right)^{2 \g}-1\right]\ 
\left\{\f{{ 
\vert xy\vert}^2}{\vert  xz\vert^2 \vert z y \vert^2} +\f{1}{\vert  
xz\vert^2} 
\f{\log \left(\f{\vert z y \vert^2}{\vert  xz\vert^2}\right)}{\log (\vert  
xz\vert^2 
\Lambda^2)}+\f{1}{\vert z y \vert^2} \f{\log \left(\f{\vert  xz\vert^2}{\vert 
z y 
\vert^2}\right)}{\log(\vert z y \vert^2 
\Lambda^2)}\right\}  \,\label{balitsky3} \ .\ea
The first term in braces corresponds to the LL kernel \eqref{eq:bfkl_ll}. The 
two other terms give an additional contribution to higher orders of 
perturbation
which by symmetry reads
\ba
 2  \int_{\cal{R}} \f{d^2 z}{ 2\pi} 
\left[ \left( \f{\vert  xz\vert}{\vert  xy\vert} \right)^{2\g}  +\left( 
\f{\vert z
 y \vert}{\vert  xy\vert} \right)^{2\g}  -1 \right]\ \f{1}{\vert  xz\vert^2} 
\f{\log 
\left({\vert z y \vert^2}/{\vert  xz\vert^2}\right)}{\log (\vert  xz\vert^2 
\Lambda^2)}\ .
\label{balitsky4}
\ea
In order to regularize \eqref{balitsky4}, we choose to introduce a truncation 
of 
the
perturbative expansion as a function of $\abar(\vert  xy\vert) \equiv 
[-b\log(\vert 
 xy\vert^2 \Lambda^2)]^{-1}.$ We will discuss later the regularization 
dependence.

Using a rescaling variable ${\bf \lambda} = {(\bf x \!-\! z})/{\vert x 
y\vert}$ and 
going to 
the complex $\lambda$-plane
one writes
\bb
& &\int_{\cal{R}} \f{d \lambda d \bar{\lambda}}{ 2 i\pi} 
\left[ \left( \lambda \bar{\lambda}\right)^{\g}  +\left( (1-\lambda)(1- 
\bar{\lambda})\right)^{\g}  -1 \right]\ \f{1}{\lambda  \bar{\lambda}} \f{\log 
\left(\f{(1-\lambda)(1- \bar{\lambda})}{\lambda \bar{\lambda}}\right)}{\log 
(\lambda \bar{\lambda}\vert x y\vert^2 \Lambda^2 )}\nn
&=& \int_{\cal{R}} \f{d \lambda d \bar{\lambda}}{ 2 i\pi} 
\left[ \left( \lambda \bar{\lambda}\right)^{\g}  +\left( (1-\lambda)(1- 
\bar{\lambda})\right)^{\g}  -1 \right]\ \f{1}{\lambda  \bar{\lambda}} \f{\log 
\left(\f{(1-\lambda)(1- \bar{\lambda})}{\lambda \bar{\lambda}}\right)}{\log 
(\lambda \bar{\lambda}) -[b \abar(\vert x y\vert)]^{-1} } \nn
&=& \int \f{d \lambda d \bar{\lambda}}{ 2 i\pi} 
\left[ \left( \lambda \bar{\lambda}\right)^{\g}  +\left( (1-\lambda)(1- 
\bar{\lambda})\right)^{\g}  -1 \right]\ \f{1}{\lambda  \bar{\lambda}} \log 
\left(\f{\lambda \bar{\lambda}}{(1-\lambda )( 1-\bar{\lambda})}\right) 
\sum_{n=1}^{N} (b \abar(\vert x y\vert))^{n}\ \log (\lambda 
\bar{\lambda}))^{n-1} 
\ ,
\label{balitsky5}\eea
where $N$ defines the level of truncation. For instance, $N=1$ corresponds 
to the 
 contribution of the running coupling to the NLL
term of the  kernel. Note that the only remaining scale dependence comes from 
$\abar(\vert x y\vert).$

It is then possible to express analytically  the expansion 
\eqref{balitsky5} term-by-term. 
One writes
\ba
\chi^{Bal}_{\cal{R}}(\g,\abar(\vert  xy\vert))  =  \chi(\g) + \abar(\vert x 
y\vert) \lim_{\e,\delta \to 0} \sum_{n=1}^{N} \left(-b 
\abar(\vert x y\vert) 
\f{\d}{\d \e}\right)^{n-1} \left(\f{\d}{\d \delta} -\f{\d}{\d \e} \right) 
I_1(\g,\e,\delta)\ ,   \label{kernelBal1} 
\ea
where the integral $I_1(\g,\e,\delta)$ is given in Appendix \ref{AI}.

In particular, the contribution ($N=1$) to the NLL  kernel reads 
\bb
\chi^{Bal}_{\cal{R}}(\g,\abar(\vert  xy\vert))  &=&   \chi(\g) + b 
\abar(\vert x y\vert) \left\{  \f{1}{2}  
\left(\chi(\g)^2+\Psi'(\g)-\Psi'(1-\g)  
\right) -\f{2 \chi(\g)}{\g}  \right\} \nn
&=&    \chi(\g) + \f{1}{\tilde{L}} \left\{  \f{1}{2}  
\left(\chi(\g)^2+\Psi'(\g)-\Psi'(1-\g)  
\right) -\f{2 \chi(\g)}{\g}  \right\} 
\,\label{kernelBal3} ,
\eea
which agrees with the result of \cite{Balitsky:2006wa}, confirmed 
by 
\cite{Kovchegov:2006wf}
(up to the inclusion of the factor $\f 53$ in a redefinition of $\Lambda$ and a 
change $\g  \to 1\!-\!\g$ w.r.t. \cite{Kovchegov:2006wf}).

Finally, the saddle-point \eqref{balitsky1} gives, in double-Mellin in 
position 
space at 
NLL order 
\ba
\omega = \f{1}{b\tilde{L}} \chi (\g) + \f{1}{b\tilde{L}^2} \chi^{(1)} 
(\g)\ ,
\label{NLL}
\ea
where 
\ba \chi^{(1)} (\g) \equiv \f{1}{2}  
\left(\chi(\g)^2+\Psi'(\g)-\Psi'(1-\g)  
\right) -\f{2 \chi(\g)}{\g}\ .
\label{NLL1} \ea

As well-known, the NLL contribution \eqref{NLL1} to the kernel has a spurious 
singularity in $1/\g^2.$ As we shall now see, this will not influence the 
determination of the universal terms in the traveling-wave solution. 
However, as explained in the introduction, it may delay the convergence of 
the asymptotic expansion.

%%%%%%%%%%%%%%%%%%%%%%%%%%%%%%%%%%%%%%%%%%%%%%%%%%%%%%%%%%%%%%%%%%%%%%%%%%%%

\subsection {Derivation of the traveling-wave solution in position space}

Keeping in a first stage the truncation at NLL level, namely  $N=1$ in 
(\ref{balitsky5},\ref{kernelBal1}), the nonlinear equation in position 
space can be formally written
\be
\label{alpha(Ltilde)}
\f 1{\bar\alpha(\tilde L)}\, \partial_Y \cN (\Lt, Y) = 
\chi_1^{Bal} (-\partial_{\Lt},\abar(\Lt))\ \cN (\Lt, Y) - \cN^{\otimes2} 
(\Lt, 
Y)\ ,
\ee
where  $\Lt = -\log [(\vert xy\vert\Lambda)^2)].$ 
One should care about the exact meaning of Eq. \eqref{alpha(Ltilde)}, which 
remains at this stage formal in two aspects. The nonlinear term  
$\cN^{\otimes2}$ 
coming  from the quadratic part 
of the integrant in \eqref{balitsky2}  is not explicitely given and the 
operator 
$\chi_1^{Bal} (-\partial_{\Lt},\abar(\Lt))$ is to be well-defined, since it 
contains two non-comutative variables.

Concerning the nonlinear terms, we know from general 
 traveling-wave properties that its precise analytic form is irrelevant 
for the asymptotic solutions. The same remains true with  
the yet unknown
NLL contributions to the LL nonlinear term.

For  the action of the kernel in Eq.\eqref{alpha(Ltilde)} we consider the  
kernel 
obtained in diagonalized form (\ref{NLL},\ref{NLL1}), namely
\ba
\label{chipos} 
\chi_1^{Bal}(\g,\abar(\Lt))= \chi(\g) + b\abar(\Lt)\ 
\chi^{(1)}(\g)\ .
\ea 
 By solving  \eqref{NLL} as a function of $\om,$ one obtains a new 
$\om$-dependent kernel $\kappa_1(\g,\omega)$ which leads to the equation
\bb
 b \tilde{L} \omega &=& \kappa_1(\g,\omega)\equiv\f{\chi(\g) + 
\sqrt{\chi(\g)^2+ 4 b \omega \chi^{(1)}  
(\g)}}{2} \ ,
\label{implicit}
\eea
where we have only kept the physical root. 
Then Eq.\eqref{alpha(Ltilde)} takes the operator form\be
\label{alpha(operator)}
b\tilde L\, \partial_Y \cN (\Lt, Y) = 
\kappa_1 (-\partial_{\Lt},\partial_{Y}) \cN (\Lt, Y) - \cN^{\otimes2} 
(\Lt, 
Y)\ .
\ee
The derivation of the traveling-wave solutions  comes from a simple extension 
of the arguments of subsection
\ref{sec:omega} on the extension of the traveling-wave method to a 
$\om$-dependent kernel. Indeed, Eq.\eqref{alpha(operator)} is  similar  to 
\eqref{alpha(L)w} in  section 
\ref{sec:alpha(L)},  substituting $L$ by $\Lt,$ see \eqref{eq:Ltilde}.

Taking into account the same  same 
substitution $L\to \Lt$ 
and considering  formula \eqref{implicit}, the traveling-wave solution is 
driven by the value of the $\om$-dependent kernel $\kappa_1(\g,\omega)$ at 
$\om = 0,$ namely
\ba
\kappa_1(\g,\omega=0)\equiv \chi(\g)\ .
\label{kern}
\ea
We see from Eq.\eqref{kern}, that  for $\omega$ small enough, that is for the 
universal terms dominant at large enough rapidity, the NLL contribution will 
not modify the saturation saddle-point $\g_c.$

From this general argument, we see that the traveling-wave solution is given 
from the LL kernel with running coupling constant in {\it position space}. 
By the momentum-position substitution property and the independence of the 
result on the precise form the nonlinear damping (despite the fact that they 
possess a different form in both representations)  one is  led to the 
universal form of the solution
\be
{\cal N}(\Lt, Y) = 
{\rm const} \cdot Y^{\frac{1}{6}} \cdot {\rm Ai} \left(
-2\log{\left[\vert xy\vert Q_s(\sqrt Y)\right]}\left(\frac{\sqrt{2 \gammac b 
\, 
\chi(\gammac)}}{\chi''(\gammac)}\right)^{\frac{1}{3}} \; Y^{-\frac{1}{6}} +
\xi_1 \right) \cdot
\left({\vert xy\vert Q_s(\sqrt Y )}\right)^{2\gammac}\ ,
\label{gluonlt}\ee
where $\xi_1, \g_c $ are the same as in   formula \eqref{gluon} (note that 
\eqref{gluonlt}, as well as \eqref{gluon}, is not valid in the infra-red. 
However,  
the infra-red region  is also universally constrained by unitarity in position 
space, 
namely ${\cal N}\sim 1$).
The saturation scale is then
 identical to \eqref{satur}
 up to a non universal multiplicative constant, meaning that the saturation 
intercept is predicted to be the same as \eqref{satur} at large rapidity.
 
\subsection{Truncation and  regularization independence} 

Let us sketch the general argument leading to the independence of the 
traveling-wave solution w.r.t. the truncation and the regularization scheme. 
For this sake, we shall use a method of ``$\om$-expansion'' introduced in the 
derivation of the linear BFKL equations at NLL order with 
renormalization-group constraints 
\cite{Ciafaloni:1998iv,Ciafaloni:1999yw,Ciafaloni:2003rd}. Briefly, this 
amounts to transform the perturbative expansion of the kernel into an 
equivalent expansion in $\om.$ 

The asymptotic universality property can be extended to a given finite value 
of $N$ in a 
more general truncation of 
the expansions (\ref{balitsky5},\ref{kernelBal1}).  Indeed  one 
can write the corresponding saddle-point equation 
as
\be
\label{eq:consistency2}
\omega = \alpha(\Lt)\ \chi_N^{Bal}(\gamma,\alpha(\Lt))\ ,
\ee
where $\chi_N^{Bal}(\gamma,\alpha(\Lt))$ is a polynomial of degree $N$ in 
$\alpha(\Lt).$ This can be considered as as an  implicit equation for  
$\Lt$ as a function of $\omega$ namely 
\be
\label{eq:implicit}
\Lt=\frac{1}{b\omega}\ \kappa_N (\gamma,\omega)\ ,
\ee
where the function $\kappa_N$ verifies thus
\be
\label{eq:implicit2}
 \kappa_N (\gamma,\omega)= 
\chi_N^{Bal}\left\{\gamma,{\omega}/{\kappa_N(\gamma,\omega)}\right\} \ .
\ee
Then, coming back to the truncation equation \eqref{kernelBal1}, one finds 
that $\chi_N^{Bal}(\g,\abar =0)= \chi(\g).$  Then the implicit equation 
\eqref{eq:implicit2} leads to $\kappa_N (\gamma,\omega\!=\!0) \equiv 
\chi(\gamma).$ One again falls in the same universality class. Indeed,  this 
is 
 expected  from the remark that $\omega$ and $\alpha(\Lt)$ are of 
the same 
order and jointly go to zero at the limit.

The same result should apply  for any consistent scheme for the  
regularization of the Landau poles, \emph{e.g.} by freezing of the coupling 
constants beyond some scale $\vert xz\vert ,\vert yz\vert >r_0$ in the 
integral \eqref{balitsky2}. The perturbative consistency of the 
regularization implies the existence of a convergence domain in  $\abar, \om$ 
near zero. In  physical terms, the traveling-wave solutions are valid in  the 
 domain  of high enough rapidity to have a perturbative  saturation scale  
and driven by the ultra-violet behaviour. Hence the nonperturbative effects 
related to the regularization should not be relevant. This is a natural 
constraint on the regularization scheme.

The regularization, which  corresponds to a given kernel $\chi^{Bal}_{\cal 
R}$ will then give rise to an  implicit equation
 \be
\label{eq:implicit3}
 \kappa_{\cal R} (\gamma,\omega)= 
\chi^{Bal}_{\cal R}\left\{\gamma,{\omega}/{\kappa_{\cal 
R}(\gamma,\omega)}\right\} \ ,
\ee
defining a new $\om$-dependent kernel $\kappa_{\cal R}.$ By the same
arguments, one again expects $\kappa_{\cal R}(\g,\om=0) \equiv \chi(\g).$

%%%%%%%%%%%%%%%%%%%%%%%%%%%%%%%%%%%%%%%%%%%%%%%%%%%%%%%%%%%%%%%%%%%%%%%%%%%%

\subsection{Traveling-wave solution in momentum space}

Now we shall demonstrate that the traveling-wave solution obtained in 
position space in the preceeding subsection is in fact in the same 
universality class as the solution found for the initial BK equation 
\eqref{alpha(L)} with running coupling in momentum space. Since a Fourier 
transform 
of the asymptotic position-space solution \eqref{gluon} is not mathematically 
justified, it is  convenient to start directly by transforming the 
equations from position to momentum space and then find the asymptotic 
solutions.

Due to the previous universality properties in this space, the problem 
reduces to the solution of the initial equation \eqref{balitsky1} when 
keeping only the LL kernel: 
\ba
b \tilde{L} \  \f{\d }{\d Y} \mathcal{N}(\vert xy\vert,Y) =  \int \f{d^2 
{z}}{2 
\pi} 
\f{\vert xy\vert^2}{\vert xz\vert ^2\ \vert zy\vert ^2}\ \left[ 
\mathcal{N}(\vert {xz}\vert ,Y) +\mathcal{N}(\vert {zy}\vert ,Y) 
-\mathcal{N}(\vert xy\vert,Y)-\mathcal{N}(\vert xz\vert ,Y)\ 
\mathcal{N}(\vert 
zy\vert ,Y) \right]\ ,
\label{balitsky6}
\ea
where the coupling depends only on the parent-dipole size $\vert xy\vert.$

Then, performing the Fourier transform \eqref{fourier} on the two sides of 
the equation gives
\ba 
\int_0^\infty \f{d \vert xy\vert}{ \vert xy\vert} \, \mathrm{J}_0 (k\vert 
xy\vert) \, b \tilde{L}\ \f{\d }{\d Y} 
\mathcal{N}(\vert xy\vert,Y) = \chi (- \d_L) N(k,Y) - N^2(k,Y) \, . \ea
Using the identity 
\ba
b \tilde{L}(\vert xy\vert) = b L(k) - b \log (\vert xy\vert^2 k^2)\ ,
\label{joli}
\ea one 
arrives at the equation:
\ba 
b L \f{\d}{\d Y} N(k, Y) - \int_0^\infty \f{d \vert xy\vert}{ \vert xy\vert} 
\, \mathrm{J}_0 (kr) 
\, 
b 
\log(\vert xy\vert^2k^2) \f{\d }{\d Y} \mathcal{N}(\vert xy\vert,Y) = \chi (- 
\d_L) N(k,Y) - N^2(k,Y) \, 
, 
\label{BKkrccB1}
\ea
where an extra term appears with respect to \eqref{alpha(L)}, 
depending still on the derivative of the position-space dipole amplitude 
$\mathcal{N}.$

Denoting $\hat{\mathcal{N}}$ the Mellin transform of $\mathcal{N}(\vert 
{xy}\vert ,Y)$ and 
$\hat{N},$  the Mellin transform of $N(k,Y),$ they can be 
simply related (see appendix 
\ref{AMellin}) by
\ba 
\hat{N}(\gamma, Y) = 2^{2 \gamma-1} \f{\Gamma (\gamma)}{\Gamma (1-\gamma)} 
\, 
\hat{\mathcal{N}}(\gamma, Y)\; .
\label{relation}\ea
For the derivative term, by insertion of Mellin-transform, one 
writes the relations
\bb & & b  \int_0^\infty \f{d \vert xy\vert}{ \vert xy\vert} \, \mathrm{J}_0 
(k\vert xy\vert) \, \log(\vert xy\vert^2k^2) \f{\d 
}{\d 
Y} \int \f{d \gamma}{2 \pi i} \, e^{- \gamma \tilde{L}} \, 
\hat{\mathcal{N}}(\gamma, Y) \ = \nn
&=& b \int \f{d \gamma}{2 \pi i} \, \f{\d }{\d Y} \hat{\mathcal{N}}(\gamma, 
Y)
\, \Lambda^{2 \gamma} \, \int_0^\infty d \vert xy\vert \, \mathrm{J}_0 
(k\vert xy\vert) \, \log(\vert xy\vert^2k^2) 
\ \vert xy\vert^{2\gamma -1}\ = \nn
&=&  b \int \f{d \gamma}{2 \pi i} \, \f{\d }{\d Y} \hat{\mathcal{N}}(\gamma, 
Y)
\, \left(\f{\Lambda}{k}\right)^{2 \gamma}\, \f{\d}{\d \gamma} \int_0^\infty d 
u 
\, \mathrm{J}_0 (u) \,  
u^{2\gamma -1}\ = \nn
 &=&  b \int \f{d \gamma}{2 \pi i} \, \f{\d }{\d Y} 
\hat{\mathcal{N}}(\gamma, Y)
\, e^{-\g L}\,
\f{\d}{\d \gamma} \left[
2^{2 \gamma-1} \f{\Gamma (\gamma)}{\Gamma (1-\gamma)}\right]\ .
\label{term}
\eea
Using now the relation \eqref{relation} between Mellin transforms, one 
finally finds the simple expression
\ba
\label{extra}
 b \int \f{d \gamma}{2 \pi i} \, e^{- \gamma L}  \ \varphi 
(\g)\f{\d }{\d Y} \hat{N}(\gamma, Y) \ ,
\ea
with $\varphi (\g)\equiv \f{\d}{\d \gamma} \log\left[
2^{2 \gamma-1} \f{\Gamma (\gamma)}{\Gamma 
(1-\gamma)}\right]=\Psi(\gamma)+\Psi(1-\gamma) + 2 \log 2$.
Hence, (\ref{BKkrccB1}) becomes 
\ba b L \f{\d}{\d Y} N(k, Y) = \chi (- \d_L) N(k,Y) + b \varphi (- \d_L) 
\f{\d 
}{\d 
Y}  N(k, Y)  - N^2(k,Y) \, . \label{BKkrccB2}\ea
Following the same derivation as  previously, the linear solution of the 
linear regime writes 
\be
\label{eq:N1}
\cN(L,Y) = \int \frac{d\gamma}{2\pi i}
         \int \frac{d\omega}{2\pi i} \,
         \cN_0(\gamma,\omega) \,
         \exp\left(-\gamma L + \omega Y + \frac{1}{b\omega} X(\gamma) + 
\int^{\g}_{\hat \g} d\g' \varphi(\g')
            \right)\ ,
\ee
where $X(\gamma)$ is the function considered in the LL case, see 
\eqref{eq:capital_x}. This solution can thus be identified  as a change of 
the kernel eigenvalue 
$\chi(\g)\to \chi(\g,\omega)\equiv \chi(\g)+b\omega\varphi(\g).$ Hence,  the 
universality 
class of the non-linear equation \eqref{BKkrccB2} is determined by 
$\chi(\g,\omega=0)=\chi(\g).$
This result can also be obtained by noting that the term 
$\int^{\g}_{\hat \g} d\g' \varphi(\g')= \log\left[2^{2 \gamma-2{\hat \g}}\ 
\f{\Gamma (\gamma)\Gamma (1-\hat \gamma)}
{\Gamma (\hat \gamma)\Gamma (1-\gamma)}\right]$ can be absorbed in a 
redefinition 
of  the impact 
factor 
$\cN_0(\gamma,\omega).$ As easy to realize from the initial saddle point 
equation \eqref{eq:saddle}, a contribution of the impact factor  may only  
consist in a change of the non universal reference  
scale $Q_0.$

The final conclusion is thus that the traveling-wave solutions of the 
Balitsky equation \eqref{balitsky1} are in the same 
universality class as the original BK equation \eqref{alpha(L)} with LL 
kernel both in 
position and in momentum space. Note that this result extends in a non trivial 
way to  running coupling a property which was easy to derive when the coupling 
is 
fixed and thus factorized from the Fourier transform.

%%%%%%%%%%%%%%%%%%%%%%%%%%%%%%%%%%%%%%%%%%%%%%%%%%%%%%%%%%%%%%%%%%%%%%%%%%%%
%%%%%%%%%%%%%%%%%%%%%%%%%%%%%%%%%%%%%%%%%%%%%%%%%%%%%%%%%%%%%%%%%%%%%%%%%%%%
%%%%%%%%%%%%%%%%%%%%%%%%%%%%%%%%%%%%%%%%%%%%%%%%%%%%%%%%%%%%%%%%%%%%%%%%%%%%

\section{Triumvirate of running couplings}
\label{Triumvirate}
Let us now apply the traveling-wave method to the evolution equation 
\eqref{EqTrium}. 
As previously, one has to consider first its linear 
part. As in section \ref{sec:alpha(r)}, one has 
to 
introduce a regularization procedure to avoid the Landau poles, except that 
the integration is now done in momentum space. Let us choose the same type of 
truncation $\cal{R}$ as for the previous equation: we expand the Landau 
poles 
denominators and truncate the resulting series.

Starting with the linear part of Eq.\eqref{EqTrium}
\bb
\f{\d }{\d Y} \tilde{\phi}(k,Y) &=&  \int_{\cal{R}} \f{d^2 \bf{q}}{2 \pi} 
\ \f{\abar 
(q^2) 
\abar (({\bf k-q})^2)}{\abar (k^2)}
 \left[\f{1}{({\bf k-q})^2} \tilde{\phi}(q,Y) + \f{1}{q^2} 
\tilde{\phi}(|{\bf 
k- q}|,Y)
 - \f{k^2}{q^2 ({\bf k-q})^2}  \tilde{\phi}(k,Y) \right] \ 
,\label{EqTriumlin}
\eea
we rewrite the triumvirate of running coupling 
\bb
\f{\abar (q^2)\ \abar (({\bf k\!-\!q})^2)}{\abar (k^2)} &=& \abar (k^2)\  
\f{\log^2(\f{k^2}{\Lambda^2})}{\log(\f{q^2}{\Lambda^2}) \, 
\log (\f{({\bf k\!-\!q})^2}{\Lambda^2}) } 
=  \f{\abar (k^2)}{\left[ 1+ b \abar (k^2) \log(\f{q^2}{k^2}) \right] \, 
\left[ 1+ b \abar (k^2) \log (\f{({\bf k-q})^2}{k^2}) \right]}
 \ .\label{Trium}
\eea
Inserting \eqref{Trium} and the Mellin representation $\tilde{\phi} 
(k,Y) = \int  \f{d \gamma}{2 \pi i} \, e^{- \gamma L} \, 
{\hat{\phi}}(\gamma,Y)$ 
in the right-hand side of \eqref{EqTriumlin}, one  obtains the diagonalized 
form of the regularized kernel $\chi^{Tri}_{\cal{R}}$
\ba
\chi^{Tri}_{\cal{R}}(\g,\abar(k^2))  =  \int_{\cal{R}} \f{d^2 \bf{q}}{2 \pi} 
\ 
\left\{\f{ \f{1}{({\bf k-q})^2} \left(\f{q^2}{k^2} \right)^{-\gamma} + 
\f{1}{q^2} \left(\f{({\bf k-q})^2}{k^2} \right)^{-\gamma}
- \f{k^2}{q^2 ({\bf k-q})^2}   }{\left( 1+ b \abar (k^2) 
\log(\f{q^2}{k^2}) \right) \, \left( 1+ b \abar (k^2) \log (\f{({\bf 
k-q})^2}{k^2}) 
\right)}\right\} \ . \ea
Using a rescaling ${\bf \lambda} = {\bf q}/{k}$ and 
going to 
the complex ${\bf \lambda}$ plane,
one writes

\bb
\chi^{Tri}_{\cal{R}}(\g,\abar(k^2)) &=& \int_{\cal{R}} \f{d \lambda d 
\bar{\lambda}}{4 \pi i}\  
\f{1}{\left[ 1+ b \abar (k^2) \log(\lambda  \bar{\lambda}) \right] \, 
\left[ 1+ b \abar (k^2) \log ((1-\lambda)(1-\bar{\lambda})) \right]} \, 
 \times  \nn
& & \quad \quad \quad \times \left[\f{\left(\lambda  \bar{\lambda} 
\right)^{-\gamma}}{(1-\lambda)\left(1-
\bar{\lambda}\right)} + \f{\left((1-\lambda)(1-\bar{\lambda}) 
\right)^{-\gamma}}{\left(\lambda  
\bar{\lambda}\right)}
- \f{1}{\left(\lambda  \bar{\lambda} (1-\lambda)(1-\bar{\lambda})\right)}   
\right] \nn
&=& \sum_{n = 0}^{N} \sum_{m = 0}^{N-n} (b \abar (k^2))^{n+m} \int \f{d 
\lambda 
d 
\bar{\lambda}}{4 \pi i} 
\log^n(\lambda  \bar{\lambda}) \, \log^m ((1-\lambda)(1-\bar{\lambda})) \, 
\left[\f{\left(\lambda  \bar{\lambda} \right)^{-\gamma}}{(1-\lambda)\left(1-
\bar{\lambda}\right)} \right. \ +\nn
& &  \quad \quad \quad \left.  +\ \f{\left((1-\lambda)(1-\bar{\lambda}) 
\right)^{-\gamma}}{\left(\lambda  
\bar{\lambda}\right)} 
- \f{1}{\left(\lambda  \bar{\lambda} (1-\lambda)(1-\bar{\lambda})\right)}   
\right] \label{KerTrium1} \ ,
\eea
where $N$ defines the level of truncation. For instance, $N\!=\!0$ 
corresponds 
to the LL BFKL kernel, and $N\!=\!1$ to the NLL
term of the kernel. It is then possible to express  
the expansion \eqref{KerTrium1} analytically term-by-term. 
One formally writes
\bb
\chi^{Tri}_{\cal{R}}(\g,\abar(k^2)) &=& \sum_{n = 0}^{N} \sum_{m = 0}^{N-n} 
(b 
\abar (k^2))^{n+m} \lim_{\e, \delta \to 0} \f{\d^n }{\d \e^n} 
\, 
\f{\d^m }{\d \delta^m} I_2 (\gamma, \e, \delta) \ ,\label{KerTrium2}
\eea
where the generating functional $I_2(\g,\e,\delta)$ is given in Appendix 
\ref{AI2}.

At the NLL level one finds 
\bb
\chi^{Tri}_{\cal{R}} (\gamma, \abar (k^2)) &=& \chi(\gamma) + b \abar (k^2) 
\left[ 
\f{\chi(\gamma)^2}{2} + \f{3}{2} \left(\Psi' (\gamma) - \Psi' (1\!-\!\gamma)  
\right) \right] + \mathcal{O}\left(\left[b \abar (k^2)\right]^2\right) 
\label{KerTrium} \ .
\eea
 Once going from 
the 
function $\tilde{\phi}$ (see \eqref{phi}) to the unintegrated gluon 
distribution 
$\phi,$ which results in  changing the factor $3/2$ into $1/2,$ this result 
gives back  the one found in \cite{Kovchegov:2006wf}.

Eq.\eqref{EqTrium} then reads in double Mellin space at 
NLL order 
\ba
\omega = \f{1}{b L} \chi (\g) + \f{1}{b L^2}\ \chi^{Tri \  (1)} 
(\g) + 
\mathcal{O}\left(\f{1}{L^3}\right)\ ,
\label{NLLTrium}
\ea
where 
\ba \chi^{Tri \  (1)}(\g) = \f{\chi(\gamma)^2}{2} + \f{3}{2} 
\left(\Psi' (\gamma) - \Psi' (1-\gamma)  \right)\ .
\label{NLLTrium1} \ea
We note that this result is different than for the dipole amplitude, 
\eqref{NLL1}, as discussed in \cite{Kovchegov:2006wf}.

Keeping in a first stage the truncation at NLL level, namely  $N\!=\!1$ in 
(\ref{KerTrium1},\ref{KerTrium2}), the nonlinear equation in momentum 
space can be formally written
\be
\label{alpha(L1)}
\f 1{\bar\alpha( L)}\, \partial_Y N (L, Y) = 
\chi^{Tri}_{1} (-\partial_L,\abar(L)) N (L, Y) - N^{\otimes2} (L, 
Y)\ ,
\ee
where
\ba
\label{chimo} 
\chi^{Tri}_{1}(-\partial_{L},\abar(L))= \chi(-\partial_{L}) + b\abar(L)\ 
\chi^{Tri \  (1)}(-\partial_{L})
\ea
with the nonlinear contribution in momentum space denoted by $N^{\otimes2} 
(L, 
Y).$ Again and also in previous QCD traveling wave studies at nonzero  
transverse momentum \cite{marquet} the precise form 
of  
$N^{\otimes2}$ will not matter for the universal behaviour of asymptotic 
solutions and thus need not be explicitely derived.

As in the previous section, one has  to solve 
\eqref{NLLTrium} as a second-order equation for $L$. One finds
\bb b L \omega &=& \f{\chi(\g) + \sqrt{\chi(\g)^2+ 4 b \omega 
\chi^{Tri \  (1)} (\g)}}{2}\ .  \eea
As explained in the introduction, and valid already for the Balitsky 
scheme, the spurious collinear pole contained in the NLL term of \eqref{chimo} 
will 
not change the universal terms, while it may delay the convergence of the 
asymptotic expansion.

The whole derivation of traveling wave solutions parallels the one described 
in 
the previous section, except for the transformation 
from position to momentum space. Hence finally the traveling wave solutions 
are 
directly expressed as in (\ref{gluon},\ref{satur}). 

Once again, the generalisation to a fixed perturbative truncation is expected 
to 
hold. The universality class of a regularized  BK 
equation with the triumvirate of running couplings is the same as the 
equation 
including the  running coupling with the ``external'' gluon momentum 
\eqref{alpha(L)}. Thus, including the effects of the transition to saturation 
leads to an unification of the asymptotic solutions, irrespective of the 
differences at the level of the solutions of the linear regime.

%%%%%%%%%%%%%%%%%%%%%%%%%%%%%%%%%%%%%%%%%%%%%%%%%%%%%%%%%%%%%%%%%%%%%%%%%%%%
%%%%%%%%%%%%%%%%%%%%%%%%%%%%%%%%%%%%%%%%%%%%%%%%%%%%%%%%%%%%%%%%%%%%%%%%%%%%
%%%%%%%%%%%%%%%%%%%%%%%%%%%%%%%%%%%%%%%%%%%%%%%%%%%%%%%%%%%%%%%%%%%%%%%%%%%%

\section{Renormalization-group improved scheme dependence}
\label{Scheme}
Let us recall why the solutions of the BFKL equation with NLL
corrections require some special treatment and why this introduces 
a NLL scheme dependence. Indeed, although the NLL 
corrections are known 
\cite{Fadin:1998py, Ciafaloni:1998gs}
they turn out to be negative and  large.
 This is due 
to spurious singularities which are in contradiction with constraints coming 
from the QCD renormalization-group properties. Indeed  the cancellation of 
these 
singularities can be obtained by suitable   contributions at higher orders. 
However, since these contributions are not yet calculated, 
there is a need for convenient schemes ensuring the compatibility with the 
renormalization group, which are not uniquely  defined. If one admits that 
the 
linear part of the various nonlinear 
evolution equations we are studying will be driven by the corresponding BFKL 
kernels, we have to take into account this scheme dependence in our 
discussion 
of universality properties. A first discussion of the scheme dependence
can be found in ref.\cite{Peschanski:2006bm}, where different schemes were 
discussed for the kernel of equation \eqref{alpha(L)}. The aim of the present 
section is to extend the discussion to the equations considered in the 
previous sections and to consider the  running coupling also in the scheme 
definition. Note that the RG-improvement plays already an important role in the
convergence aspects of the universality properties, see section II-C.

For our discussion, one could consider at least two different classes of 
schemes, following Ref.  \cite{Peschanski:2006bm}. In a first class, 
containing \emph{e.g.} the schemes of Refs.
\cite{Ciafaloni:1999yw,Ciafaloni:2003rd}, the dependence of the scheme on the 
higher orders of the perturbation expansion has been expressed through the 
$\om$-dependence, following the method of \cite{Ciafaloni:1998iv}. In this 
case, the higher order 
resummation appears in the  kernel only through the  dependence over the two 
Mellin variables, \emph{i.e.} $\chi (\gamma \!=\! -\partial_L, \omega 
\!=\!\partial_Y),$ 
where  $\omega \!= \!{\cal O}(\bar\alpha)$ drives the higher-order 
corrections.
For this class of two-variable NLL kernels, the general argument that the 
traveling-wave solutions are driven by $\chi(\g,\omega\!=0\!)=\chi(\g)$ 
applies, see  section II-C.

In a  
second class of schemes such as S3 and S4 \cite{Salam:1998tj}, the analysis 
of  \cite{Peschanski:2006bm} leads to  results depending 
on 
the value chosen for the fixed coupling appearing in the definition of the 
renormalization-group 
scheme. Indeed, the traveling-wave solutions are driven by the kernel $\chi 
(\gamma, \omega \!=\!0, \bar\alpha \ne 0) \ne \chi (\gamma).$ Hence, even if 
the 
form of the solutions is similar \cite{Peschanski:2006bm}, the critical 
parameters are different and thus  the traveling-wave solutions   appear to 
be  scheme-dependent.  One practical question, for instance, is to know what 
are the predictions for the traveling-wave solutions when choosing the 
coupling defining the scheme  varying with the momentum scale, see 
\emph{e.g.} \cite{Peschanski:2004vw}. Indeed, one would prefer a unified 
treatment of the running coupling 
effects taking into account the running 
coupling in the RG improved kernel itself. We shall show, that 
in this case, and on variance with the results at fixed coupling for the 
regularization, the same universality class defined by the LL kernel (with 
running 
coupling) is recovered. 

For this sake, we can consider  in the same way all previous  
equations, either that corresponding to the BK 
equation in transverse momentum, \eqref{eq:saddle}, or in transverse
 position space \eqref{NLL}, or for the case 
of the triumvirate  \eqref{NLLTrium}, using  any of the renormalization-group 
improved  NLL kernels. Then the corresponding saddle-point equations become 
\be
\label{eq:consistency3}
\omega = \alpha(L\ or\ \Lt)\ \chi^{NLL}(\gamma,\omega,\alpha(L\ or\ \Lt))\ .
\ee
Now, in parallel with the previous discussion, Eq.\eqref{eq:consistency3} can 
be interpreted in all cases as an implicit 
equation for  $L$ or $\Lt$ namely 
\be
\label{eq:implicitn}
L\ or\ \Lt=\frac{1}{b\omega}\ \kappa^{NLL} (\gamma,\omega)\ ,
\ee
where the function $\kappa^{NLL}$ verifies the implicit equation
\be
\label{eq:implicit2n}
 \kappa^{NLL} (\gamma,\omega)= 
\chi^{NLL}\left\{\gamma,\omega,\f{\omega}{\kappa^{NLL}(\gamma,\omega)}
\right\} 
\ 
.
\ee
This boils down to reformulate  schemes, starting from   S3 and S4, by using  
 an appropriate $\om$-expansion \cite{Ciafaloni:1998iv}. These schemes are 
now in the same first class discussed above. 

Then, the condition  $\kappa^{NLL}(\gamma,\omega\!=\!0) \equiv \chi(\gamma)$ 
is expected to be verified in all cases by perturbative consistency, and thus 
by the same argument, one again falls in the same universality class. Indeed, 
if we use the $\om$-expansion method to redefine the   schemes S3 or S4, it 
is possible to 
show that $\kappa^{NLL} 
(\gamma,\omega\!=\!0)\!=\!\chi(\gamma,\omega\!=\!0,\alpha\!=\!0) 
\!=\!\chi(\gamma).$ All other schemes verifying this relation will again fall 
in the same  universality class, defined by the traveling-wave solutions 
(\ref{gluon},\ref{satur}). The question remains for further study whether 
other types of NLL schemes, such as the one of  Ref.\cite{Brodsky:1998kn} can 
be given a similar 
treatment.

%%%%%%%%%%%%%%%%%%%%%%%%%%%%%%%%%%%%%%%%%%%%%%%%%%%%%%%%%%%%%%%%%%%%%%%%%%%%
%%%%%%%%%%%%%%%%%%%%%%%%%%%%%%%%%%%%%%%%%%%%%%%%%%%%%%%%%%%%%%%%%%%%%%%%%%%%
%%%%%%%%%%%%%%%%%%%%%%%%%%%%%%%%%%%%%%%%%%%%%%%%%%%%%%%%%%%%%%%%%%%%%%%%%%%%

\section{Summary and predictions}
\label{Conclusion}
To summarize, we considered three versions of the QCD evolution equation in 
rapidity in the 
mean-field approximation, \emph{i.e.} the Balitsky-Kovchegov 
\cite{Balitsky:1995ub,Kovchegov:1999yj,Kovchegov:1999ua} equation \eqref{1}, 
 extended to take into account the 
running coupling of QCD. The first one considers the 1-dimensional BK 
equation for the dipole 
amplitude in momentum space,  with the substitution of the fixed coupling by 
a running coupling in terms of the gluon transverse momentum, see 
\eqref{alpha(L)}. The two other forms come from recent theoretical 
calculation of the  quark-loop contributions. They differ by a separation 
scheme for higher orders between  the running coupling  and the kernel.
The 
second equation we consider is written  for the dipole amplitude in 
transverse position space , see \eqref{balitsky1} \cite{Balitsky:2006wa}.
 The 
last one can also be written for the dipole amplitude in transverse position 
space, leading  to a ``triumvirate'' of running 
couplings, and differs from the previous one. We study it in its equivalent 
formulation as the evolution equation for the unintegrated  gluon 
distribution in 
momentum space see \eqref{EqTrium} \cite{Kovchegov:2006vj,Kovchegov:2006wf}. 
We 
have 
also enlarged the discussion by considering the modified BK equations 
including the 
renormalization-group 
improvements of the NLL BFKL kernels 
\cite{Salam:1998tj,Ciafaloni:1999yw,Ciafaloni:2003rd}. 

Let us summarize our results. The saturation effects which are formulated 
through the nonlinear damping 
terms in the BK equation leads to asymptotic traveling-wave solutions for all 
equations. Their behaviour at 
high rapidity is highly universal. More precisely the form of the solution at 
high rapidity   and the two first terms of  the rapidity expansion of the 
 the saturation intercept ${d\log(Q_s)}/{dY}$ are the same. Hence they are 
identical  to  those found and derived in
\cite{Munier:2003sj} for  the  initial BK equation with running coupling 
\eqref{alpha(L)}.

For establishing this strong universality property, We have examined the 
different types of higher order dependence which modify the solutions of the 
linear regime
\bi
\ii {\it Observable dependence:}\ 
The  traveling wave solutions at 
high rapidity happen to be  independent of the  considered distribution 
functions, by 
contrast with the 
solutions of the linear part of the equations. For instance the dipole 
amplitude 
and the unintegrated gluon distribution which lead to different rapidity 
dependence in the linear regime for the same scheme \cite{Kovchegov:2006wf} 
have the same saturation intercept and traveling-wave
spectrum. This leads to an unified asymptotic NLL predictions for observables, 
independently of its formulation in terms of dipole or gluon distributions.
\ii {\it Separation scheme dependence:}\  By computing the traveling-wave 
solutions for both the Balitsky scheme \cite{Balitsky:2006wa} and the 
Kovchegov-Weigert one \cite{Kovchegov:2006vj,Kovchegov:2006wf}, we find the 
same results, belonging to the universality class of the simpler 
equation\eqref{alpha(L)} . 
\ii {\it Regularization dependence:}\ 
The definition of equations (\ref{balitsky1},\ref{EqTrium}) 
requires 
a regularization, since their consistent formulation should avoid Landau-pole 
singularities. 
Using a finite truncation of the QCD perturbative expansion, we show that the 
order of truncation, which affects the linear regime, does not change 
the asymptotic traveling-wave solutions. We propose a general argument for 
general perturbatively consistent regularization schemes.
\ii {\it RG-improved scheme dependence:}\  The QCD coupling appears also in 
the definition 
of appropriate schemes for 
using the NLL BFKL kernels in order to avoid spurious singularities. The 
improved NLL kernels with constant coupling were shown to give 
scheme-dependent traveling wave solutions \cite{Enberg:2006aq}, even if the 
overall coupling is  running \cite{Peschanski:2006bm}. If the scheme is 
defined using the running coupling through the $\om$-expansion 
\cite{Ciafaloni:1998iv}, we recover the strong universality property.  
\ii {\it Position \emph{vs.} momentum-space dependence:}\ The universality 
class
is the same when the equations are expressed either in position, or 
momentum-space formalism by Fourier transformation. This shows that the 
saturation intercept is the same. The form of the front is also invariant 
except 
for the infra-red (with respect to the saturation scale) regions which are not 
bound by universality properties. We have already mentionned that the infra-red 
region in both representations is also universally constrained,  by unitarity, 
namely 
${\cal N}\sim 1$ in position space  and  $N \sim \log(Qs/k)$ in momentum.
 
\ii {\it Rapid \emph{vs.} slow convergence:}\ By discussing the effects of 
the singularities contained in the NLL kernels, we have distinguished 
between the RG-improved kernels and those containing spurious singularities. Indeed, the 
rapidity region where the universal properties appear is expected to be 
delayed in the case where such singularities remain. It is thus important 
to use, whenever known, a RG-improved scheme for exhibiting  the 
universality properties.
\ei

From our results, we conclude that the saturation effects on the rapidity 
evolution of the dipole amplitude or the unintegrated gluon distribution  
with running coupling give 
a stabilisation of the solutions with respect to higher order 
corrections at high enough rapidity. They appear to be independent 
of higher order contribution 
either to the linear kernel or to the nonlinear damping terms. however, subasymptotic
 nonuniversal contributions may be present.

In mathematical terms one expresses this property as the existence of a large 
 {\it 
universality class} of solutions given by (\ref{gluon},\ref{satur}).

At this stage, it is possible to formulate predictions for the behaviour of 
the exact solutions of the evolution equations which can  be obtained 
through numerical simulations. We can list them as follows:
\bi
\ii The form of the traveling-wave fronts (except  in the full saturation 
region) should converge at high rapidity for the solution of the different 
equations in the same representation space, either in position or momentum.
\ii The saturation-scale intercept ${d\log(Q_s)}/{dY}$ which can be 
determined from these fronts should converge to the same function.
\ii Geometric scaling  in $\sqrt Y,$ related to the constant limit of the 
wave speed should be seen from the solutions, at least in the  middle of the 
front.
 \ei
These predictions can be tested using numerical simulations of the rapidity 
evolution corresponding to  modified BK equations with running coupling. In 
Ref.\cite{albacete1}, the running coupling was heuristically introduced for  
models in position space and compared with the fixed coupling case. 
Naturally, they could not include the recent theoretical advances, but some 
of the conclusions are still interesting to quote and seem to fit  
approximately with our  predictions, even if the equations are different. The 
traveling-wave regime is observed at high enough rapidity, the saturation 
scales for running coupling seem to converge, at variance with the fixed 
coupling case where they depend on the value of the coupling. The form of the 
front verifies geometric scaling.

While completing this theoretical analysis we were informed of interesting 
numerical simulations of the relevant equations \cite{albacete} using a 
regularization scheme based on the freezing of the coupling. The simulations 
\cite{albacete} seem to indicate that the traveling-wave 
structure is preserved at moderate rapidities and independent of the initial 
conditions. However, the wave parameters seem to differ from the critical ones. 
We predict that 
at higher enough 
rapidity this difference will decrease and  disappear. Note that  arguments 
based on subasymptotic   
parametric solutions of the nonlinear 
equations  \cite{Peschanski:2005ic} show  that the traveling wave structure 
can be maintained at 
moderate rapidity but with 
modified speed as observed in \cite{albacete}. Also the form of the wave front 
is modified, which could give an explanation of the difference already noticed 
in \cite{albacete1} between the running and non running cases.

On a phenomenological ground, such a theoretical solution is characterized by 
a 
geometric scaling \cite{geom} in $\sqrt Y,$ either in momentum or position 
space. 
By contrast with  fixed coupling which leads to geometric scaling in $ Y$ 
\cite{Mueller:2002zm,Munier:2003vc}, this property is characteristic of the 
BK 
equation with running coupling 
\cite{Triantafyllopoulos:2002nz,Munier:2003sj}. 
Geometric scaling in $ Y$ or $\sqrt Y$  seem to be consistent with data on 
the 
proton structure functions \cite{Gelis:2006bs}. However, the precise 
comparison 
with the parameters of the theoretical traveling-wave solutions show that 
nonuniversal terms are present, and thus nonasymptotic contributions cannot 
be neglected. Parametric solutions, as in \cite{Peschanski:2005ic} could help 
to take into account nonuniversal contributions.

As an outlook for further directions of theoretical study, we  mention  the 
extension of our investigation beyond the mean-field approximation 
leading to the BK equation. For this sake, one has to consider the  hierarchy 
of 
QCD evolution equations including  fluctuations, {\it i.e.} Pomeron-loop 
terms, 
when the coupling constant is running. Also the role of impact factors, 
\emph{e.g.} for the coupling to the virtual photon in deep inelastic 
scattering 
has not been yet studied at NLL level. Let us also note that a third (and 
last) universal term in the asymptotic expansion of the saturation intercept, 
which 
is known for the 
fixed coupling case \cite{Munier:2003vc},  could be the  remaining track of   
the NLL kernels in the universal traveling-wave solutions as discussed after 
equation \eqref{eq:bk_Lt}. Finally a specific study of subasymptotic terms is 
deserved. It may lead to traveling-wave properties at more moderate rapidities 
and thus interesting on a phenomenological point of view.

\begin{acknowledgments}
We thank   Cyrille Marquet and Sebastian Sapeta for fruitful discussions at the 
origin at the present work. We are grateful to   Ian Balitsky and Yuri 
Kovchegov 
for useful explanations on their 
recent papers and Javier Albacete for valuable information on very recent 
results with  Yuri Kovchegov before publication. 
\end{acknowledgments}

%%%%%%%%%%%%%%%%%%%%%%%%%%%%%%%%%%%%%%%%%%%%%%%%%%%%%%%%%%%%%%%%%%%%%%%%%%%%
%%%%%%%%%%%%%%%%%%%%%%%%%%%%%%%%%%%%%%%%%%%%%%%%%%%%%%%%%%%%%%%%%%%%%%%%%%%%
%%%%%%%%%%%%%%%%%%%%%%%%%%%%%%%%%%%%%%%%%%%%%%%%%%%%%%%%%%%%%%%%%%%%%%%%%%%%

\appendix

\section{Calculation of $I_1(\g,\e,\delta)$} \label{AI}

The integral to be computed in \eqref{balitsky5} is
\bb  \!  \! \! \! \! \! \! \! \! \! \! \! \! \!& & I_1(\g,\e,\delta) \equiv   
\int 
\f{d \lambda d \bar{\lambda}}{ 2 i\pi} 
\left[ \left( \lambda \bar{\lambda}\right)^{\g}  +\left( (1-\lambda)(1- 
\bar{\lambda})\right)^{\g}  -1 \right]\ \f{1}{(\lambda  
\bar{\lambda})^{1+\e} 
(1-\lambda)^{\delta}(1- \bar{\lambda})^{\delta}}\nn
 \!  \! \! \! \! \! \! \! \! \! \! \! \! \! &=&\f{\Gamma (\g-\e) \Gamma 
(1-\delta) \Gamma (-\g + \e +\delta)}{\Gamma 
(1-\g+\e) \Gamma (\delta) \Gamma (1+\g - \e -\delta)} + \f{\Gamma (-\e) 
\Gamma 
(1+\g-\delta) \Gamma (-\g + \e +\delta)}{\Gamma (1+\e) \Gamma (-\g+\delta) 
\Gamma (1+\g - \e -\delta)} - \f{\Gamma (-\e) \Gamma (1-\delta) \Gamma ( \e 
+\delta)}{\Gamma (1+\e) 
\Gamma (\delta) \Gamma (1 - \e -\delta)} \, .\eea
Then, one finds
\bb \lim_{\delta \to 0}\left(\f{\d}{\d \delta}\! -\!\f{\d}{\d \e} \right) 
I_1(\g,\e,\delta)&=&
\f{1}{\e^2}   \!-\!\f{1}{(\g-\e)^2} \! -\!\f{\g^2}{\e (\g-\e)^2} J(\g,\e) 
\left[\f{1}{\e} \!+\!\Psi(1-\e)\! +\! 
\Psi(1+\e) \!-\! \Psi (\g)\! - \!\Psi(1-\g) \!-\! \f{2}{\g}  \right]
\label{result}\eea
where 
\ba J(\g,\e) \equiv \f{\Gamma (1-\e) \Gamma (\g) \Gamma (1-\g + \e 
)}{\Gamma 
(1+\e) \Gamma (\g-\e) \Gamma (1-\g)}
= 1 - \e \chi(\g) + \f{\e^2}{2} [\chi(\g)^2 + \Psi'(1-\g)-\Psi'(\g)] + 
\mathcal{O}(\e^3) \, .
\ea
Applying the expansion operator \eqref{kernelBal1} on the result 
\eqref{result}, 
one writes
\ba
b \abar(r^2) \lim_{\e \to 0} \sum_{n=1}^{N} \left(-b \abar(r^2) 
\f{\d}{\d \e}\right)^{n-1} \left\{\f{1}{\e^2}   \!-\!\f{1}{(\g-\e)^2} \! 
-\!\f{\g^2}{\e (\g-\e)^2} J(\g,\e) 
\left[\f{1}{\e} \!+\!\Psi(1-\e)\! +\! 
\Psi(1+\e) \!-\! \Psi (\g)\! - \!\Psi(1-\g) \!-\! \f{2}{\g}  \right]\right\}
\,\label{kernelBal4} ,
\ea
At first next-to-leading order, one finds
\bb
& & b \abar(r^2) \lim_{\e \to 0} \sum_{n=1}^{N} \left(-b \abar(r^2) 
\f{\d}{\d \e}\right)^{n-1} \left\{  \f{1}{2}  
\left(\chi(\g)^2+\Psi'(\g)-\Psi'(1-\g)  \right) -\f{2 \chi(\g)}{\g} + 
\mathcal{O}(\e) \right\}\nn
&=&  b \abar(r^2) \left\{  \f{1}{2}  \left(\chi(\g)^2+\Psi'(\g)-\Psi'(1-\g)  
\right) -\f{2 \chi(\g)}{\g}  \right\} + \mathcal{O}(b^2 \abar^2(r^2)) \nn
&=&  \f{1}{\tilde{L}} \left\{  \f{1}{2}  
\left(\chi(\g)^2+\Psi'(\g)-\Psi'(1-\g)  
\right) -\f{2 \chi(\g)}{\g}  \right\} + 
\mathcal{O}\left(\f{1}{\tilde{L}^2}\right) 
\,\label{kernelBal5} .
\eea

%%%%%%%%%%%%%%%%%%%%%%%%%%%%%%%%%%%%%%%%%%%%%%%%%%%%%%%%%%%%%%%%%%%%%%%%%%%%

\section{Mellin-transforms in position \emph{vs.} momentum} 
\label{AMellin}

The Mellin-transforms of the dipole amplitude are defined  in 
position space as
\ba \mathcal{N}(r,Y) = \int \f{d \gamma}{2 \pi i} \, e^{- \gamma \tilde{L}} 
\, 
\hat{\mathcal{N}}(\gamma, Y)
= \int \f{d \gamma}{2 \pi i} \int \f{d \omega}{2 \pi i} \, e^{\omega Y - 
\gamma 
\tilde{L}} \, \tilde{\mathcal{N}}(\gamma, \omega)\ , \label{Melr2}
\ea
and in momentum space as
\ba N(k,Y) = \int \f{d \gamma}{2 \pi i} \, e^{- \gamma L} \, \hat{N}(\gamma, 
Y)
= \int \f{d \gamma}{2 \pi i} \int \f{d \omega}{2 \pi i} \, e^{\omega Y - 
\gamma 
L} 
\, \tilde{N}(\gamma, \omega)\ . \label{Melk}
\ea
Let us  relate these Mellin-transformed amplitudes by  rewriting the initial 
dipole amplitude in momentum 
space \eqref{fourier}.
\bb N( k,Y) &=&  \int_0^\infty \f{d r}{ r} \, \mathrm{J}_0 (kr) \, \int \f{d 
\gamma}{2 \pi i} \, e^{- \gamma \tilde{L}} \, \hat{\mathcal{N}}(\gamma, Y)
=  \int \f{d \gamma}{2 \pi i}\, \hat{\mathcal{N}}(\gamma, Y) \,   \Lambda^{2 
\gamma} \, \int_0^\infty d r \, \mathrm{J}_0 (kr) \ r^{2 \gamma-1}\nn
&=&  \int \f{d \gamma}{2 \pi i}\, \hat{\mathcal{N}}(\gamma, Y) \,   
\Lambda^{2 
\gamma} \, k^{-2 \gamma} 2^{2 \gamma-1} \f{\Gamma (\gamma)}{\Gamma 
(1-\gamma)}=  
\int \f{d \gamma}{2 \pi i} \, e^{- \gamma L} \, 2^{2 \gamma-1} \f{\Gamma 
(\gamma)}{\Gamma (1-\gamma)} \, \hat{\mathcal{N}}(\gamma, Y)\; .
\label{final}\eea
Hence, by comparing \eqref{Melk} with the last expression in 
Eq.\eqref{final}, 
the two  Mellin transformed functions are found related by
\ba \hat{N}(\gamma, Y) = 2^{2 \gamma-1} \f{\Gamma (\gamma)}{\Gamma 
(1-\gamma)} 
\, 
\hat{\mathcal{N}}(\gamma, Y)\; .
\label{relation1}\ea

%%%%%%%%%%%%%%%%%%%%%%%%%%%%%%%%%%%%%%%%%%%%%%%%%%%%%%%%%%%%%%%%%%%%%%%%%%%

\section{calculation of $I_2(\g,\e,\delta)$} \label{AI2}

The integral to be computed in \eqref{KerTrium2} is
\bb \!\!\!\!\!\!\!\!\!\!\!\!I_2 (\gamma, \e, \delta) &=&  \int \f{d \lambda d 
\bar{\lambda}}{4 \pi 
i} \, 
\left(\lambda  \bar{\lambda} \right)^{\e}  
\left((1\!-\!\lambda)(1\!-\!\bar{\lambda}) 
\right)^{\delta} \, \left[\f{\left(\lambda  \bar{\lambda} 
\right)^{-\gamma}}{(1\!-\!
\lambda)\left(1\!-\!\bar{\lambda}\right)}  \!+\! 
\f{\left((1\!-\!\lambda)(1\!-\!\bar{\lambda}) 
\right)^{-\gamma}}{\left(\lambda  \bar{\lambda}\right)} 
- \f{1}{\left(\lambda  \bar{\lambda} 
(1\!-\!\lambda)(1\!-\!\bar{\lambda})\right)}   
\right] \nn
\!\!\!\!\!\!\!\!\!\!\!\!\!\!\!\!\!\!\!\!&=& \f{\Gamma(1-\gamma+\e) 
\Gamma(\delta)\Gamma(\gamma\!-\!\e\!-\!\delta)}{2 
\Gamma(\gamma\!-\!
\e) \Gamma(1\!-\!\delta) \Gamma(1\!-\!\gamma\!+\!\e\!+\!\delta)} \!+\! 
\f{\Gamma(\e) 
\Gamma(1\!-\!\gamma\!+\!
\delta)\Gamma(\gamma\!-\!\e\!-\!\delta)}{2 \Gamma(1\!-\!\e) 
\Gamma(\gamma\!-\!\delta) 
\Gamma(1\!-\!
\gamma\!+\!\e\!+\!\delta)} \!-\! \f{\Gamma(\e) 
\Gamma(\delta)\Gamma(1\!-\!\e\!-\!\delta)}{2 
\Gamma(1\!-\!\e) 
\Gamma(1\!-\!\delta) \Gamma(\e\!+\!\delta)} 
\  . \label{ITrium}
\eea
The next-to-leading result comes from the expansion
\ba
 I_2 (\gamma, \e, \delta) = \chi(\gamma) +(\e+\delta) \left[ 
\f{\chi(\gamma)^2}{4} + \f{3}{4} 
\left(\Psi' (\gamma) - \Psi' (1-\gamma)  \right) \right] + 
\mathcal{O}(\e^2)+ 
\mathcal{O}(\e \delta)+ \mathcal{O}(\delta^2)\ .
\ea
%%%%%%%%%%%%%%%%%%%%%%%%%%%%%%%%%%%%%%%%%%%%%%%%%%%%%%%%%%%%%%%%%%%%%%%%%%%%
%%%%%%%%%%%%%%%%%%%%%%%%%%%%%%%%%%%%%%%%%%%%%%%%%%%%%%%%%%%%%%%%%%%%%%%%%%%%
%%%%%%%%%%%%%%%%%%%%%%%%%%%%%%%%%%%%%%%%%%%%%%%%%%%%%%%%%%%%%%%%%%%%%%%%%%%%

%%%%%%%%%%%%%%%%%%%%%%%%%%%%%%%%%%%%%%%%%%%%%%%%%%%
%\bibliographystyle{h-physrev3}
%\bibliography{sat}

\end{document}